\documentclass[12pt,a4paper]{article}

\usepackage{latexsym,amssymb,lastpage}
\usepackage{graphicx,amsfonts}
\usepackage{times,mathptmx,bm}
\usepackage{dcolumn}
\usepackage{stmaryrd}
\usepackage{mathdots,amsbsy}
\usepackage{amsmath}
\usepackage[utf8]{inputenc}

\def \bm#1{\mbox{\boldmath{$#1$}}}   
\newcommand{\ii}{\textrm{i}}

\usepackage{geometry} 

\usepackage{natbib}

\numberwithin{equation}{section}

\begin{document}
\title{Torsion instability of soft solid cylinders}
\author{Pasquale Ciarletta  \\[2pt]
CNRS and Universit\'{e} Pierre et Marie Curie - Paris 6,\\ [0pt]
Institut Jean le Rond d'Alembert, UMR 7190,\\
4 place Jussieu case 162, 75005 Paris, France, \\[24pt]
Michel Destrade\\[2pt]
NUI Galway\\
School of Mathematics, Statistics and Applied Mathematics,\\ [0pt]
University Road, Galway, Ireland. \\[24pt]
\textit{ This article is dedicated to Ray Ogden in friendship and esteem} }

\date{}
 \maketitle
 


\begin{abstract}
{The application of a pure torsion to a long and thin cylindrical rod is known to
provoke a twisting instability, evolving from an initial kink to a
knot.
In the torsional parallel-plate rheometry of short and stubby cylinders, the geometrical
constraints impose zero-displacements of the axis of the cylinder,
preventing the occurrence of such a twisting instability.
Under these experimental conditions, wrinkles occur on the cylinder's surface at a given critical angle of torsion.
Here we investigate this subclass of elastic instability -- which we call \textit{torsion instability} -- of soft cylinders subject to a combined finite axial stretch and torsion, by applying the theory of incremental elastic deformation superimposed on finite strains.
We formulate the incremental boundary elastic problem in the Stroh differential form, and use the surface
impedance method to build a robust numerical procedure for deriving the marginal stability curves.
We present the results for a Mooney-Rivlin material and study the influence of the material parameters on the
elastic bifurcation.}
\\[24pt]

\end{abstract}

\newpage


\section{Introduction}
\label{intro}


The application of combined finite axial stretch and finite torsion to a solid right
cylinder constitutes one of the few universal solutions of nonlinear
isotropic incompressible elasticity, where here ``universal'' means
that the deformation can be achieved for any homogeneous
hyperelastic material by the application of surface tractions alone.

The simple torsion of a solid cylinder can be defined as the
deformation by which planes perpendicular to the axis of the
cylinder are rotated in their own plane through an angle
proportional to their distance from one end surface. In a seminal
paper, \citet{rivlin48} found that a state of simple torsion can be
maintained by surface tractions alone (end couples and end
compressive normal forces) for any incompressible, neo-Hookean
material. Further extensions were made in subsequent papers of the
same series by Rivlin, who calculated analytical expressions of such
surface tractions for a generic incompressible, isotropic material
\citep{rivlin48b}, and for a hyperelastic tube subjected to combined
axial stretching and torsion \citep{rivlin49}. In that latter paper,
Rivlin mentions that Dr H.A. Daynes had drawn his attention to the
earlier work of \citet{poynting}, who had observed and measured the
lengthening of a steel wire and of a rubber rod upon twisting, and
Rivlin provided a satisfactory theoretical explanation to this
phenomenon, later referred to as the \textit{positive Poynting
effect}.
The results of Rivlin were revisited and extended by many over the
years. For instance, \citet{horgan99} found that, if the strain
energy density function used to model the behaviour of the cylinder
depended only on the first principal strain invariant, then there
must exist a universal relation between the surface force and the
torque (universal relative to the class of incompressible materials
with strain energy depending only on the first principal invariant).
Such a relative-universal relation is unlikely to be observed in
practice, indicating that the strain energy should also depend on the
second invariant. Additional discussion on this subject was later
provided by \citet{wineman05}.

\begin{figure}[!ht]
\centerline{\includegraphics[height=5cm]{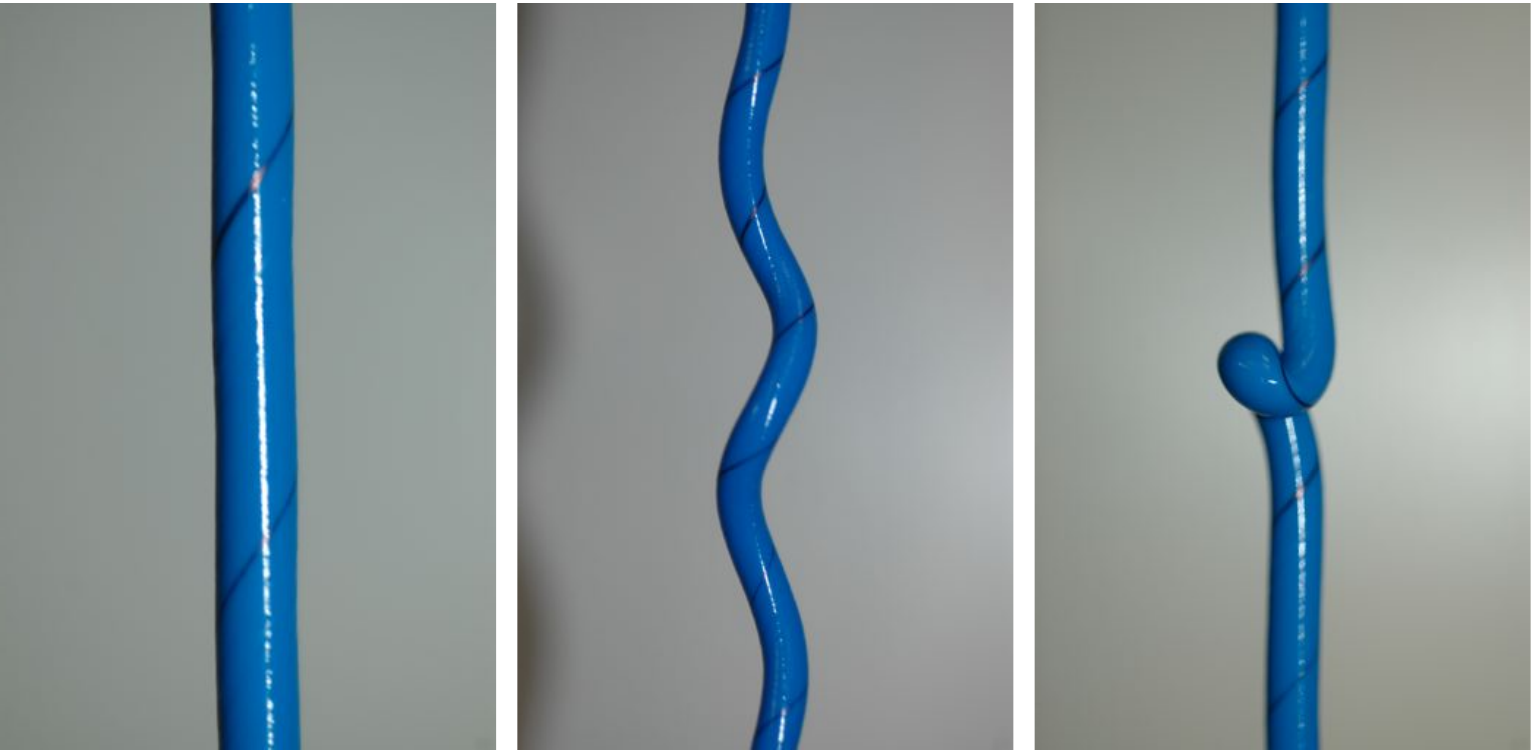}}
\caption{Twisting a long cylindrical rod made of polyurethane: at
first a large torsion takes place, followed by ``twisting
instability'' allowing for displacements on the axis, eventually
turning into a twisting knot.} \label{fig1}
\end{figure}

It is intuitive to expect that the application of a compressive
axial force during simple torsion should eventually lead to a
buckling instability once a certain threshold of torsion rate is
reached. However there are very few studies of \emph{torsional
instabilities} on a solid cylinder to be found in the literature.
\citet{green} found an analytical solution for neo-Hookean solids in
the subclass of instability modes giving rise to finite
displacements on the axis of the cylinder. \cite{duka93}  later
studied the numerical solution of this instability subclass for a
generic Mooney-Rivlin solid. As initially guessed by  \citet{green},
their solution represents the ``twisting instability'' of a
cylindrical rod, evolving from an initial kink to a knot, see Figure
\ref{fig1}, and \citet{gent} later presented an energetic analysis
of this transition.

\begin{figure}[!ht]
\centerline{\includegraphics[height=5cm]{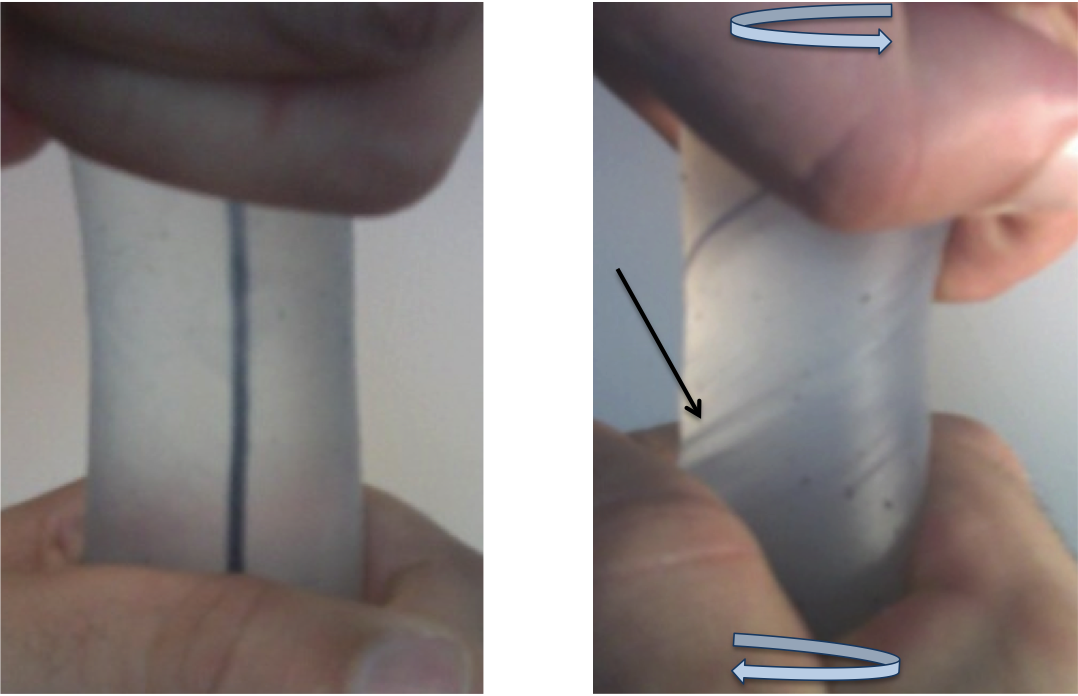}}
\caption{A short cylinder made at silicone rubber at rest (left) and
after applying by hands a torque at the end surfaces (right). We
marked a solid black line to follow the deformation on the cylinder,
while the arrow indicates the wrinkles formed at a critical torsion
rate.} \label{fig1bis}
\end{figure}

A recent experiment by \citet{mora11}  shows that when a cylindrical
sample of soft gel with small axial length/external radius ratio
is deformed by a rheometer, it eventually displays a wrinkling
instability pattern on its surface at a finite critical value of the
torsion rate, see their Figure 4. In that experimental scenario,
the possibility of a displacement on the axis of the cylinder is
prevented by the geometrical constraints, and so an instability
other than twisting is observed. This particular effect can be easily reproduced by applying a torque by hands on a soft, short cylinder, as reproduced in Figure \ref{fig1bis}.
The aim of our work is to
investigate the onset of this new subclass of wrinkling
instabilities for an isotropic, incompressible, hyperelastic
material. We call them \emph{torsion instabilities}.

The paper is organised as follows. In Section \ref{sec2}, we recall
the equations governing the kinematics of the finite axial
stretching and torsion  of the soft cylinder,  and derive the
axial-symmetric  solution. We specialise the analysis to the class
of Mooney-Rivlin materials in order to keep the number of
constitutive parameters low, while retaining dependence of the
strain energy density on the first two principal strain invariants.
In Section 3, we  perform an incremental (linearised) stability
analysis, by superposing a small-amplitude perturbation on this
basic finite deformation. The incremental boundary-value problem is
derived and the corresponding numerical results are presented in
Section 4 and discussed in Section 5.


\section{Finite torsion and stretching of a soft elastic cylinder}
\label{sec2}


Let us consider an  elastic cylinder made of an isotropic, homogeneous, nonlinearly elastic, incompressible material,
with axial length $L$ and external radius $R_o$ in the fixed
reference configuration $\Omega_0$.
Using cylindrical coordinate systems, the kinematics of the deformation can be defined by a mapping $\pmb{\chi}$: $\Omega_0\rightarrow \Re^3$ that brings the
material point ${\bf X} = {\bf X}(R,\Theta,Z)$ to the spatial
position ${\bf x} = {\bf x}(r,\theta,z)= \pmb\chi({\bf X})$ in the
deformed configuration, where $(R, \Theta, Z)$ and $(r,\theta,z$) are the coordinates along the orthonormal vector bases $(\bf E_1, \bf E_2, \bf E_3)$ and $(\bf e_1, \bf e_2, \bf e_3)$, respectively.
In particular, the soft cylinder is subjected to a finite stretching and torsion, so that:
\begin{equation}
r= \dfrac{R}{\sqrt{\lambda_z}}, \qquad
\theta=\Theta+ \gamma \lambda_z Z, \qquad
 z= \lambda_z Z,
 \label{kin}
\end{equation}
where $\gamma$ is the torsion angle per unit length and $\lambda_z$ is the uniform
stretching ratio in the axial direction ($\gamma L$ is the angle of torsion of the whole cylinder).
In the current deformation, the cylinder has radius $r_o=R_o/\sqrt{\lambda_z}$ and length $l = \lambda_z L$.

From Eq.(\ref{kin}), the deformation gradient ${\bf F} =
\partial \pmb \chi/\partial {\bf X}$ has the following components in the $\bf e_i \otimes \bf E_j$ basis,
\begin{equation}
 {\bf F} = \begin{bmatrix}
\dfrac{1}{\sqrt{\lambda_z}}& 0& 0 \\ 0 & \quad
\dfrac{1}{\sqrt{\lambda_z}} \quad & \quad r \gamma {\lambda_z} \\ 0 &
0 & {\lambda_z} \end{bmatrix}, \label{F}
  \end{equation}
and clearly, $\det {\bf F}=1$, so that the imposed deformation is volume preserving and compatible with the constraint of incompressibility.

From a constitutive viewpoint, we assume from now on that the cylinder behaves
as a \emph{Mooney-Rivlin hyperelastic material}, so that its strain energy
density $W$ is given by
\begin{equation}
W = \dfrac{c_1}{2}  \left(I_1 -3 \right) + \dfrac{c_2}{2} \left(I_2 -3 \right),  \label{const}
\end{equation}
where $c_1$, $c_2$ are positive material constants (and $\mu = c_1 + c_2$ is the shear modulus) and $I_1,
I_2$ are the first two principal invariants of the left Cauchy-Green
deformation tensor $\mathbf b = \mathbf{FF}^T$, where the superscript $T$ denotes the transpose.
Note that not all generality is lost by specialising the strain energy so early.
In particular, the Mooney-Rivlin strain energy function encompasses all weakly non-linear elastic incompressible models up to the third-order in the strain, as shown by \citet{rivlin51} (see \citet{DeGM10} for another proof).
Also, as long as $c_2 \ne 0$, the problem of a relative-universal relation, unlikely to be observed in practice, is avoided (more on this later).

From \eqref{F}, we find the components of $\bf b$ in the $\bf e_i \otimes \bf e_j$ basis as
\begin{equation}
 {\bf b} = \begin{bmatrix}
\lambda_z^{-1} & 0&  0 \\ 0 & \quad
\lambda_z^{-1} + \gamma^2 r^2 \lambda_z^2 \quad & \quad r
\gamma {\lambda_z}^2 \\ 0 & r \gamma {\lambda_z}^2 & {\lambda_z}^2
\end{bmatrix}, \label{b}
\end{equation}
and thus compute its principal invariants as
\begin{equation}
I_1={\rm tr} \ {\bf b}=
{\lambda_z}^2+ 2 \lambda_z^{-1} + \gamma^2 r^2 \lambda_z^2,
\qquad
I_2=\tfrac{1}{2}\left[{\rm tr} \ ({\bf b}^2)-{\rm tr}^2 \
{\bf b}\right]= 2{\lambda_z}+ \lambda_z^{-2}+\gamma^2 r^2
\lambda_z^2. \label{I2}
\end{equation}

Using the constitutive relation in Eq.\eqref{const}, the Cauchy
stress tensor $\pmb \sigma$ can be written, by Rivlin's
representation theorem, as
\begin{equation}
\pmb \sigma= c_1 {\bf b} -c_2 {\bf b}^{-1} -p {\bf I}, \label{sigma}
\end{equation}
where ${\bf I}$ is the second-order identity tensor, and $p$ is a
Lagrange multiplier arising from the incompressibility constraint
$(\det {\bf F} -1)=0$, to be determined from the equations of equilibrium and the boundary conditions.
Here, the non-zero components of the
Cauchy stress from Eq.\eqref{sigma} read
\begin{equation}
\begin{array}{ll}
& \sigma_{rr}= {c_1}{\lambda_z^{-1}}-c_2 \lambda_z  -p,\\[3pt]
&\sigma_{\theta\theta}= c_1\left({\lambda_z^{-1}}+\gamma^2 r^2
\lambda_z^2\right)-c_2 \lambda_z -p,  \\[3pt]
& \sigma_{zz}=c_1 \lambda_z^2 -
c_2\left({\lambda_z^{-2}}+\gamma^2 r^2 \lambda_z\right)-p, \\[3pt]
& \sigma_{\theta z}=\sigma_{z \theta}= c_1\gamma r \lambda_z^2-c_2
{\gamma r \lambda_z},
\end{array}\label{sigmacomp}
\end{equation}
and, in the absence of body forces, the
equilibrium equations are
\begin{equation}
\text{div} \; \pmb \sigma= \bf 0. \label{div}
\end{equation}
For the axis-symmetric deformation fields in Eq.\eqref{kin},
the only non-vanishing equilibrium equation in Eq.\eqref{div} is
\begin{equation}
\frac{\partial (r \sigma_{rr})}{\partial r} -\sigma_{\theta \theta}
=0, \label{eq}
\end{equation}
and the traction-free boundary conditions at the external radius are
\begin{equation}
\sigma_{rr} (r_o) =0 \qquad {\rm at } \quad r_o= R_0/\sqrt{\lambda}_z.
\label{bc0}
\end{equation}

By integrating $\sigma_{rr}$ in Eq.\eqref{eq}, subject to Eq.\eqref{bc0}, the Lagrange multiplier $p$ can be determined as:
\begin{equation}
p= \frac{c_1}{\lambda_z}-c_2{\lambda_z}+ \int_r^{r_o} c_1 r \gamma^2
\lambda_z^2 dr =c_1 \left[\frac{1}{\lambda_z}+\frac{\gamma^2 \lambda_z^2}{2}(r_o^2-r^2)\right]-c_2{\lambda_z}.\label{p0}
\end{equation}
Finally, it is straightforward to show that such a finite torsion
and stretching of the cylinder can be  obtained by applying the following normal force $N$ and a torque $M$ on the end surfaces,
\begin{equation}
N = 2\pi \int_{r=0}^{r_o} \sigma_{zz} r dr =
 \pi{R_o}^2 \left[ (\lambda_z - \lambda_z^{-2}) (c_1+c_2\lambda_z^{-1}) - \dfrac{\gamma^2 {R_o}^2}{2}( c_1 + 2c_2 \lambda_z^{-1})  \right], \label{N}
\end{equation}
and
\begin{equation}
M=2\pi \int_{r=0}^{r_o} \sigma_{z\theta} r^2 dr = \dfrac{\pi \gamma
{R_o}^4}{2}  (c_1 + c_2 \lambda_z^{-1}),
\label{M}
\end{equation}
where Eqs.(\ref{sigmacomp}, \ref{p0}) have been used, see \citet{rivlin51}.

In \emph{simple torsion}, there is no axial stretch: $\lambda_z=1$, and those expressions simplify to
\begin{equation} \label{NM-MR}
N =   - \dfrac{\pi \gamma^2 {R_o}^4}{4}( c_1 + 2c_2), \qquad
M=\dfrac{\pi \gamma
{R_o}^4}{2}  (c_1 + c_2).
\end{equation}
Here we first notice that an axial compression is needed for
imposing simple torsion because $N<0$. This is the so-called {\it
positive Poynting effect} for hyperelastic solids, which is a
nonlinear elastic effect forcing the spread of the top surfaces of a
cylinder under torsion. This axial compression increases
quadratically with an increasing torsional strain, and it is
reasonable to hypothesise that a buckling instability may occur
beyond a certain critical torsion. We investigate this possibility
in the next section. Second, we note that if the dependence on $I_2$
was dropped by taking $c_2=0$ above (neo-Hookean case), then the
following relative-universal relation would be in force:
$2N/\gamma^2 = - M/\gamma$. In their experiments on a vulcanized
rubber cylinder of length and radius $L=R_o= 2.5$ cm, \cite{rivlin51}
found indeed that $N$ and $M$ were proportional to $\gamma^2$ and
$\gamma$, respectively, in agreement with Eq.\eqref{NM-MR}. However,
they found that  $2N/\gamma^2 = -0.0212$ N/m$^2$ and $M/\gamma =
0.0157$ N/m$^2$ \citep{drozdov}, showing that the relative-universal
relation does not hold and that the rubber in question must be
modelled as a Mooney-Rivlin material, not as neo-Hookean.


\section{Linear stability analysis}


In this section,  we perform a linear stability analysis of the
axis-symmetric solution for a soft cylinder subjected to finite
torsion and stretching. For this purpose, we first introduce the
theory of  incremental deformations superimposed on finite
strains. Next, we derive the incremental equilibrium equations and
boundary conditions in the Stroh form and finally, we propose a robust
numerical procedure for solving the resulting  boundary value
problem.


\subsection{Incremental elastic theory}


We perform a perturbation of the large
axis-symmetric solution given by Eqs.(\ref{kin}, \ref{sigmacomp},
\ref{p0}) by using the theory of incremental deformations superimposed on finite
strains \citep{ogden97}. In practice, we proceed by writing
\begin{equation}
\textbf{x}_p=\pmb \chi(\textbf{X})+ \pmb{\dot \chi}(\textbf{x})
\end{equation}
where $\textbf{x}_p$ is the perturbed spatial position, and we
assume that the incremental displacement $\pmb{\dot \chi}$ is infinitesimal and represents a first-order correction.

We define the associated spatial displacement gradient as $\rm
\pmb{\Gamma}={\rm grad} \: \bm{\dot \chi}$, and by simple
differentiation rules we find that the perturbed deformation
gradient ${\bf \dot F}_p$ reads:
\begin{equation}
{\bf F}_p={\bf F}+{\bf \dot F}={\bf F}+ { \rm \pmb\Gamma}{\bf F},
\label{deltaF}
\end{equation}
where $ {\bf \dot F}$ is the incremental deformation gradient.
Performing a series expansion of the constitutive relation in
Eq.\eqref{sigma} to the first order, we express the components of
the incremental stress $ \mathbf{ \dot S}$, i.e. the push-forward of
the nominal stress,  as:
\begin{equation}
 \dot S_{ji} = {L}_{jikl} \, \Gamma_{kl} + p\, \Gamma_{ji} - \dot p
\, \delta_{ji}, \label{S}
\end{equation}
where $\dot p$ is the increment in the Lagrange multiplier $p$,
$\delta_{ji}$ is the Kronecker delta,  and ${L}_{jikl}$ are
the components of the fourth-order tensor of instantaneous moduli,
i.e. the push-forward of the fixed reference elasticity tensor.
Explicitly,
\begin{equation}
{L}_{jikl}=F_{j\gamma}F_{k\beta}\dfrac{\partial^2
W}{\partial F_{i \gamma} \ \partial F_{l \beta} }, \label{L}
\end{equation}
where Einstein's summation rule on repeated indices is assumed.
For instance, the moduli for the Mooney-Rivlin material \eqref{const} are
\begin{equation}
L_{jilk}= c_1 b_{jl} \delta_{ik} + c_2\left[2b_{ij}b_{kl} + (b_{nn}b_{jk} - (\mathbf b^2)_{jk})\delta_{il}
 - b_{il}b_{jk} - b_{ik}b_{jl}\right].
\end{equation}

Now, the incremental equilibrium equations take the
following form:
\begin{equation}
{\rm div} \, \mathbf{\dot S} = \mathbf 0, \label{incsigma}
\end{equation}
whilst the vanishing of the incremental traction at the free surface
gives:
\begin{equation}
 \dot {S}_{rr}= \dot {S}_{r\theta}= \dot {S}_{rz}=0 \qquad
 {\rm at} \quad r_o= R_0/\sqrt{\lambda}_z.\label{incbc}
\end{equation}
Finally, the incremental incompressibility constraint is written
as:
\begin{equation}
\rm tr \ {\pmb \Gamma}=0. \label{inc}
\end{equation}
Eqs.(\ref{incsigma}-\ref{inc}) represent the incremental boundary
value problem, whose solution is now investigated.


\subsection{Stroh formulation of the incremental problem}


We develop the incremental deformation fields $ \pmb \chi = [u_r, u_\theta ,
u_z]^T$ by separation of variables, in the following form:
\begin{equation}
u_r (r, \theta ,z)= U(r) \cos(m \theta -k_z z),\qquad
 {[u_\theta (r,\theta ,z), u_z(r, \theta ,z)]^T}= [V(r), W(r)]^T \sin(m \theta -k_z
z),
\label{u}
\end{equation}
where $U$, $V$, $W$ are functions of $r$ only,  the
\emph{circumferential mode number} $m$ is an integer, and the
\emph{axial wavenumber} $k_z$ is a real number.

\begin{figure}[!ht]
\centerline{\includegraphics[height=5.2cm]{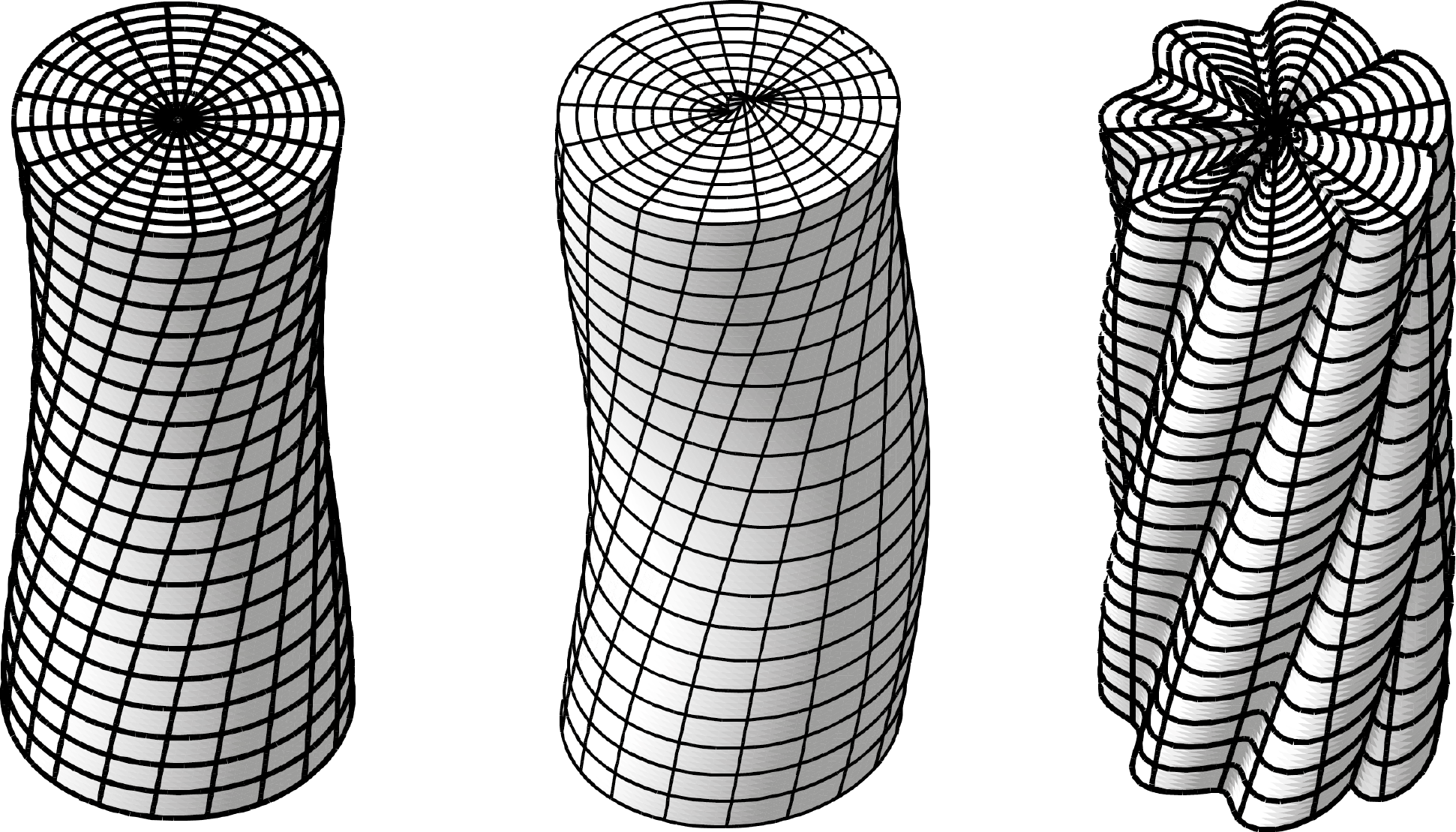}}
\caption{Surface pattern generated by a torsion instability
perturbation superimposed on the finite torsion of a right cylinder.
Here we implement a perturbation in the form of a torsion
instability where for illustrative purposes we take the angle of
torsion to be $60^\circ$, no axial pre-stretch, and the amplitude of
the perturbation to be one-tenth of the current radius. We chose to
have one wavelength axially (all three pictures) and, in turn,  a
circumferential wave number $m=0$ (left picture), $m=1$ (middle
picture), and $m=7$ (right picture).} \label{fig-torsinst}
\end{figure}

Figure \ref{fig-torsinst} displays different modes that can be expressed by such a perturbation.
For illustrative purposes there, we took a right cylinder of stubbiness $L/r_o=5$, subject to a large torsion of angle $\gamma L = 60^\circ$ and no axial pre-stretch ($\lambda_z=1$), onto which we superimposed a perturbation in the form of Eq.\eqref{u}, with $U(r_o)=V(r_o)=W(r_o)=0.1 r_o$ (small amplitude perturbation) and $k_z
r_o= 1$ (axial wave number)
(For a more accurate picture of an actual torsion instability, the critical parameters $\gamma$, $k_z$, and the relative displacements $U(r_o)$, $V(r_o)$, $W(r_o)$, must be computed from the stability analysis of Section 4, depending on the values of the constitutive parameters $c_1$ and $c_2$.)
By observation of Eq.\eqref{u} and of the figure we can confirm that from now on we can discard the case $m=0$ because it
represents an axis-symmetric perturbation, of no relevance to torsional instability.
We further notice that the pictures for the $m=0$ and $m=1$ perturbation modes closely match those of the early Euler buckling of a right cylinder under compression, see \cite{depascalis}.

We now try to model the experimental conditions at play in the torsion and stretch deformation field created by a rheometer.
Because there, the top and bottom faces are glued to  plates rotating about a fixed axis, we impose that the centre of the top face be directly aligned with that of the bottom face.
This is achieved when
\begin{equation} \label{axial-number}
k_z = \dfrac{2n\pi}{l} = \dfrac{2n\pi}{\lambda_z L},
\end{equation}
where the integer $n$ is the \emph{axial mode number}.

By substituting Eq.\eqref{u} into Eq.\eqref{S}, we can express the components of the incremental stress tensor in a  form similar to that of the displacements,
\begin{align}
& [\dot S_{rr} (r, \theta ,z),\dot p (r, \theta ,z)]^T= [ S_{rr}(r), P(r)]^T \cos(m \theta -k_z z), \notag \\
& {[\dot S_{r\theta} (r,\theta ,z), \dot S_{rz}(r, \theta ,z)]^T}= [
 S_{r\theta}(r), S_{rz}(r)]^T \sin(m \theta -k_z z), \label{s}
\end{align}
say, where $P$ and $S_{ij}$ are functions of $r$ only, and ``$\ii$'' is the imaginary unit.

This formulation allows a great
simplification of the boundary value problem.
Hence, the incompressibility
constraint in Eq.\eqref{inc} can be rewritten as:
\begin{equation}
U+m V+r (U^{'}-{k_z} W)=0,
\label{inc2}
\end{equation}
where the prime denotes differentiation with respect to $r$.
Moreover, the increment $\dot p$ of the Lagrange multiplier can be
found from the constitutive equation for $\dot{s}_{rr}$ , as
follows:
\begin{equation}
P = -r  S_{rr}+L_{rr\theta\theta} (U+mV) -k_z r
L_{rrz\theta} V+m L_{rr\theta z}W-k_z r L_{rrzz} W+r(p+L_{rrrr})
U^{'}.
\label{dp}
\end{equation}

Further simplifications arise once we introduce the \emph{displacement-traction
vector} $\pmb \eta$ as:
\begin{equation}
{\pmb \eta}= [U, V, W, \ii r{S}_{rr}, \ii r{S}_{r\theta}, \ii r {S}_{rz}]^T.
\end{equation}
Indeed, using Eqs.(\ref{u}, \ref{s},\ref{dp}), it is possible to
rewrite the entire boundary value problem given by Eqs.(\ref{S},
\ref{incsigma},\ref{inc}) as the following first-order
differential system:
\begin{equation}
\frac{\text d{\pmb \eta}}{\text dr}= \frac{\ii }{r}{\bf G}{\pmb \eta},
 \label{stroh}
\end{equation}
which is called the \textit{Stroh formulation} of the
incremental problem. In particular, the Stroh matrix ${\bf G}$
admits the following block representation:
\begin{equation}
{\bf G}= \left[\begin{array}{cc} {\bf G}_1 & {\bf G}_2\\{\bf G}_3
&{\bf G}_1^{+}
\end{array}\right],
 \label{G}
\end{equation}
where the $3 \times 3$ sub-blocks ${\bf G}_2$ and ${\bf G}_3$ are real symmetric, and the
symbol ${+}$ denotes the adjugate (i.e. transpose conjugate) matrix
operator. In particular, the matrices ${\bf G}_1$ and ${\bf G}_2$
admit the following simplified representations,
\begin{equation}
{\bf G}_1= \left[\begin{array}{ccc} \ii & \ii  m & - \ii  k_z r\\[3pt]
\dfrac{\ii  \delta_1}{\beta} & -\dfrac{\ii \delta_2}{\beta} &0\\[8pt]
-\dfrac{\ii  \delta_3}{\beta}& -\dfrac{\ii  \delta_4}{\beta}
&0\\
\end{array}\right], \qquad
{\bf G}_2= \left[\begin{array}{ccc} 0&0&0\\ 0&\dfrac{L_{rzrz}
}{\beta}& -\dfrac{L_{r\theta rz} }{\beta}\\[8pt]
0& -\dfrac{L_{r\theta rz} }{\beta}&\dfrac{L_{rzrz} }{\beta}
\end{array}\right],
 \label{G2}
\end{equation}
with
\begin{align}
& \beta={L_{r\theta rz}^2 -L_{r\theta r \theta}L_{rzrz}}, \notag \\
& \delta_1= m [(p+L_{r\theta \theta r}) L_{rzzr} - L_{r\theta rz} L_{rz\theta r}]
  + k_{z} r [-L_{r\theta zr}L_{rzrz} +L_{r\theta rz} (p+L_{rzzr})],
\notag \\
& \delta_2= -(p+L_{r\theta \theta r})
L_{rzrz}+L_{r\theta rz} L_{rz\theta r},
\notag \\
&  \delta_3 = m [(p+L_{r \theta e \theta}) L_{rzr\theta} - L_{r\theta r \theta} L_{rzr\theta}] + k_{z} r [-L_{r\theta z r}L_{r\theta rz} +L_{r\theta r\theta} (p+L_{rzzr})],
\notag \\
& \delta_4= (p+L_{r\theta r \theta})
L_{rzr\theta}-L_{r\theta r\theta}L_{rz\theta r}.
\end{align}
The expression of ${\bf G}_3$ is very lengthy and although we have obtained it formally with a Computer Algebra System, we do not report here for the sake of brevity.
Great simplifications arise when dealing with a neo-Hookean material ($c_1 = \mu$, $c_2 = 0$), as reported in
Appendix A.

Finally, considering that $m$ and $k_z$ are integer- and
real-valued, respectively, it is easy to show that the Stroh matrix
${\bf G}$ displays the following symmetry \citep{shuvalov03}
\begin{equation}
{\bf G}= {\bf T} {\bf G}^{+} {\bf T},
\end{equation}
where ${\bf T}$ has zero diagonal sub-block matrices, whilst
off-diagonal blocks are identity matrices.

The Hamiltonian structure and algebraic properties of the Stroh matrix form the basis of the robust asymptotic and numerical procedures presented in the next sections.
We note that \citet{duka93} have also put the incremental problem in the form of a first-order differential system, although not in the present, optimal, Stroh form.


\subsection{Numerical solution using the solid-cylinder impedance matrix}


Following \citet{shuvalov03}, we rely on the \emph{impedance matrix method}
for solving the incremental elastic problem.

Let us define the $6 \times 6$
matricant ${\bf M}(r, r_i)$ as the solution of the initial value
problem:
\begin{equation}
\left[\frac{\text d}{\text dr}-\frac{\ii }{r} {\bf G}(r)\right]{\bf M}(r, r_i)=
{\bf 0} \qquad {\rm with} \quad {\bf M}(r_i, r_i)={\bf I}_{(6)},
\quad r,r_i\neq0. \label{matr}
\end{equation}
Thus, from Eqs.(\ref{stroh},\ref{matr}), the displacement-traction
vector solution can be expressed as follows:
\begin{equation}
{\pmb \eta}(r)= {\bf M}(r, r_i) {\pmb \eta}(r_i)=
 \left[\begin{array}{ll}{\bf M}_1(r, r_i) & {\bf M}_2(r, r_i) \\{\bf M}_3(r, r_i) & {\bf M}_4(r, r_i)\end{array} \right]{\pmb \eta}(r_i).
 \label{eta}
\end{equation}
Now if we define:
\begin{equation}
{\bf u}= [U(r), V(r), W(r)]^T, \qquad {\bf t}= [{S}_{rr}(r),
{S}_{r\theta}(r), S_{rz}(r)]^T,
\end{equation}
as the displacement and traction vectors, respectively, then the
displacement-traction vector can be expressed as ${\pmb \eta}= [{\bf
u}, \ii  r \, {\bf t} ]^T$. It is then possible to define a
functional relation between the traction and the displacements
vectors, reading:
\begin{equation}
 r \,{\bf t} = {\bf Z} \,  {\bf u}, \label{impZ}
\end{equation}
where ${\bf Z}$ is a \textit{surface impedance matrix}. {Here, one
can build an expression for ${\bf Z}$ either by using the
\textit{conditional} (i.e depending on  its value at $r=r_i$)
\emph{impedance matrix}, as ${\bf Z}={\bf Z}(r, r_i)=-\ii{\bf
M}_3(r, r_i){\bf M}^{-1}_1(r, r_i)$ {for a stress-free boundary
condition imposing} ${\bf Z}(r_i)=0$, or by using the
\textit{solid-cylinder impedance matrix} ${\bf Z}={\bf Z}(r)$,
independent of  an auxiliary condition at some other coordinate, see
\cite{shuvalov10} for details.}

In either way, the Stroh formulation in Eq.\eqref{stroh} can be
manipulated by substituting Eq.\eqref{impZ} to eliminate the
dependence on ${\bf u}$. Accordingly, the differential system is
transformed into a \emph{differential Riccati  equation} for ${\bf Z} $,
namely
\begin{equation}
\frac{\text{d}}{\text dr}{\bf Z} =\frac{1}{r} \left({\bf G}_3 + {\bf Z} {\bf G}_2
{\bf Z} - \ii{\bf Z}{\bf G}_1 + \ii {\bf G}_1^{+} {\bf Z}\right).
\label{RiccZ}
\end{equation}

Recalling, from Eq.\eqref{M}, that the matricant solution diverges
for $r\rightarrow0$, this method is of no use for determining the
solution on the cylinder axis. Conversely, the limiting value ${\bf
Z}_0\equiv{\bf Z}(r=0)$ in Eq.(\ref{impZ}), also known as
\textit{central-impedance matrix} \citep{shuvalov10},  is of utmost importance for
solving the incremental problem of a solid cylinder.
\citet{shuvalov03} demonstrated that the fundamental solution of Eq.
\eqref{stroh}, having a regular singular point at $r=0$, can be
expressed in the form of a Frobenius series depending on the
eigenspectrum of the matrix $\ii {\bf G}(0)$. From Eqs.(\ref{F},
\ref{L}, \ref{G}, \ref{G2}), we find that the sub-blocks of ${\bf G}(0)$ read:
\begin{align}
& {\bf G}_1(0)= \left[\begin{array}{ccc} \ii  & \ii  m & 0\\
\bar\delta& \bar\delta &0\\
0&0
&0\\
\end{array}\right], \notag \\
&{\bf G}_2(0)=
\left[\begin{array}{ccc}
0&0&0\\
 0&-\dfrac{\lambda_z (c_1 \lambda_z  + c_2 )}{c_1^2\lambda_z +  c_1 c_2 \left(1+\lambda_z^3\right) + c_2^2 \lambda_z^2} &0\\
 0&0&-\dfrac{\lambda_z (c_1 \lambda_z  + c_2 )}{c_1^2\lambda_z +  c_1 c_2 \left(1+\lambda_z^3\right) + c_2^2 \lambda_z^2}
\end{array}\right] \label{G20}
\end{align}
with $\bar\delta = \ii\frac{2 c_2^2 \left(1+\lambda_z^3\right)+c_2
\lambda_z^3 \left(-\gamma^2  R_o^2+2  \lambda_z\right) c_1 +
\lambda_z^2 \left(-2-\gamma^2 R_o^2 \lambda_z^2\right) c_1^2}{2
\lambda_z \left(c_2^2 \lambda_z^2+c_2 \left(1+\lambda_z^3\right) c_1
+\lambda_z c_1^2\right)}$, and:
\begin{equation}
{\bf G}_3(0)=\left[\begin{array}{ccc}
 -\frac{\ii \left(-16-4 \gamma^2 \left(R_o^2-m^2 R_o^2\right) \lambda_z^2+\gamma^4 m^2 R_o^4 \lambda_z^4\right)}{4 \lambda_z} & -\frac{\ii m \left(-16+\gamma^4 R_o^4 \lambda_z^4\right)}{4 \lambda_z} & 0 \\[6pt]
 -\frac{\ii m \left(-16+\gamma^4 R_o^4 \lambda_z^4\right)}{4 \lambda_z} & \frac{\ii \left(4 m^2 \left(4+\gamma^2 R_o^2 \lambda_z^2\right)+\lambda_z^2 \left(-4 \gamma^2 R_o^2-\gamma^4 R_o^4 \lambda_z^2\right)\right)}{4 \lambda_z} & 0 \\[6pt]
 0 & 0 & \frac{\ii m^2}{\lambda_z}
\end{array}\right] \label{G30}
\end{equation}

Using Eqs.(\ref{G20}, \ref{G30}), it is now possible to show that the
eigenvalues of $\ii {\bf G}(0)$ are $\lambda_G= \left\{\pm(
m-1)\right.$, $\pm m$, $\left.\pm( m+1)\right\}$, all independent
on the imposed deformation and on the material properties. In
particular, we find that all such eigenvalues $\lambda_G$ differ by
an integer, so that the Frobenius power expansion requires the
introduction of additional terms compared to the solution given by
\citet{shuvalov10}.
Moreover, in the case $m=1$ we also find that $\ii{\bf G}(0)$ is not semi-simple, due to the presence of rigid-body motions.
We discuss this special case in further details in the next section.

For our purposes, it is easier to identify the central impedance matrix
${\bf Z}_0$ as the stable solution of the following algebraic Riccati
equation \citep{shuvalov10},
\begin{equation}
{\bf G}_3(0) + {\bf Z}_0 {\bf G}_2(0) {\bf Z}_0 - \ii{\bf Z}_0{\bf
G}_1(0) + \ii {\bf G}_1^{+}(0) {\bf Z}_0={\bf 0}, \label{Riccdis}
\end{equation}
hence avoiding non-physical singularity at $r=0$ in Eq.
\eqref{RiccZ}. {The stable solution is the} unique, symmetric,
semi-definite solution ${\bf Z}_0$ {which can be found} by imposing
that all eigenvalues of the matrix $ - \ii {\bf G}_1(0)+{\bf G}_2(0)
{\bf Z}_0$ be negative. Considering the Taylor expansion ${\bf
Z}(r)= \sum_{n=0}^{\infty}{\bf Z}_n r^n$, the matrix ${\bf Z}_1$ can
be calculated from Eq.\eqref{RiccZ} at the first order in $r$, as
the stable solution of the following algebraic Riccati equation:
\begin{multline}
{\bf G}_3(r)+  {\bf Z}_0 {\bf G}_2(r) {\bf Z}_0 - \ii{\bf Z}_0{\bf
G}_1(r) + \ii {\bf G}_1^{+}(r) {\bf Z}_0 +r^2 {\bf Z}_1 {\bf G}_2(r)
{\bf Z}_1\\ - \ii r {\bf Z}_1\left[{\bf G}_1(r)- {\bf G}_2(r) {\bf
Z}_0-\frac{\ii}{2}{\bf I} \right] + \ii r \left[{\bf G}_1^\dag(r)- {\bf Z}_0 {\bf
G}_2(r) +\frac{\ii}{2}{\bf I}\right]{\bf Z}_1 ={\bf 0}.
\label{Riccdis2}
\end{multline}
Choosing a starting point $r_c\ll1$, the solution of the incremental
problem can be found by  integrating numerically ${\bf Z}(r)$ in
Eq.\eqref{RiccZ} using the initial value ${\bf Z}(r_c)={\bf Z}_0+r_c
{\bf Z}_1$, with ${\bf Z}_0$ and ${\bf Z}_1$ given by
Eqs.(\ref{Riccdis}, \ref{Riccdis2}), respectively. Performing
iterations on all the coefficients determining the order parameter
of the bifurcation, the target condition of the numerical integration
is given by the boundary condition in Eq.\eqref{incbc}, and reads:
\begin{equation}
\det {\bf Z}(r_o)= {\bf 0}.\label{bcZ}
\end{equation}

In summary, the elastic boundary problem is transformed into the differential
Riccati equation for ${\bf Z}$ in Eq.\eqref{RiccZ}, which can be integrated
numerically by  imposing the non-singularity of the
central-impedance matrix ${\bf Z}_0$ and the boundary condition in
Eq.\eqref{bcZ}. The numerical results are detailed in the
next section.


\section{Results}


In this section, we present   results for the linear stability
analysis of a soft solid cylinder subjected to finite torsion and
axial stretching.
We report in turn  results for neo-Hookean materials ($c_1=\mu$, $c_2=0$) and for
Mooney-Rivlin materials ($c_1\neq0$, $c_2 \neq 0$).

As explained in the previous section, the case $m=1$ requires a
special consideration because the matrix $\ii {\bf G}(0)$ then has a
doubly degenerate eigenvalue. As a consequence, the
central-impedance matrix ${\bf Z}_0$ is semi-definite, indicating
the occurrence of rigid-body motion modes.
As first reported by
\citet{green} for a neo-Hookean solid, finite non-zero displacements
on the axis of the cylinder arise only in the case $m=1$, leading to the classical problem of a twisted Euler rod, forming an helix of pitch $1/k_z$.
This instability can evolve with the sudden
onset of a sharply bent ring, or knot (see Figure \ref{fig1}), as investigated by
\citet{gent}.
Eventually, the helical patterns can turn into localised writhing in the
post-buckling torsional behaviour \citep{thompson}, which can explain, for
example, DNA supercoiling \citep{neukirch}.
Although such effects
have been widely investigated in the  literature, there seems to be
no information to be found regarding higher order torsion instabilities.
For this reason, we leave aside the case $m=1$ in what follows, and we
investigate the onset of torsion instabilities with $m\geq 2$, corresponding
to a zero displacement on the axis of the cylinder.


\subsection{Torsion instabilities for a neo-Hookean material}


In the case of a neo-Hookean material (i.e. $c_1=\mu$, $c_2=0$), the
Stroh matrix has the simplified form given in Appendix A. As first
reported by \citet{green}, the corresponding incremental boundary
value problem admits the following analytical solution:
\begin{align}
& U(r)= \sum\limits_{j=1}^3 A_j\left[I_{m-1}(q_j r) - \dfrac{m(\lambda_z^{-3} q_j^2 + 2\gamma(m\gamma-k_z) - (\gamma - k_z)^2)}{q_j r(\lambda_z^{-3}q_j^2 -(m\gamma-k_z)^2)} {I_{m}(q_j r)}\right], \notag \\[6pt]
& V(r)=\dfrac{1}{r} \sum\limits_{j=1}^3 A_j\left[ \dfrac{2\gamma
r(m\gamma-k_z)}{\lambda_z^{-3} q_j^2 - (m\gamma-k_z)^2} I_{m-1}(q_j r) - \dfrac{m(\lambda_z^{-3} q_j^2+2\gamma(m\gamma-k_z) - (m\gamma-k_z)^2)}{q_j (\lambda_z^{-3} q_j^2 - (m\gamma-k_z)^2)} {I_{m}(q_j r)}\right], \notag \\[6pt]
& W(r)=\sum\limits_{j=1}^3 A_j\left[\dfrac{q_j}{k_z}{I_{m}(q_j r)}\right],
\label{solI}
\end{align}
where $A_j$ are arbitrary constants, $I_{m}(q_j r)$ is the modified
Bessel function of the first kind  of order $m$, and $\pm q_j$ are
the distinct roots of the following characteristic equation:
\begin{multline}
\lambda_z^{-6} q^6 -\lambda_z^{-3} \left[\lambda_z^{-3} k_z^2 + 2(m\gamma-k_z)^2 \right]q^4 \\
 + (m\gamma-k_z)^2 \left[2\lambda_z^{-3} k_z^2 + (m\gamma-k_z)^2 \right] q^2
  + k_z^2(m\gamma-k_z)^2 \left[4\gamma^2-(m\gamma-k_z)^2 \right] = 0.
\label{char}
\end{multline}
However, if the roots of Eq.\eqref{char} are not all distinct
(e.g. for $m\gamma=k_z$ there are four roots equal to zero), then the
solution in Eq.\eqref{solI} is no longer valid.
Therefore, the proposed  numerical procedure relying on the Riccati equation in
Eq.\eqref{RiccZ} with Eqs.(\ref{Riccdis}-\ref{bcZ}) is much easier and robust to run to completion in order to derive the instability threshold than by  the means of  the analytical solution.

\begin{figure}[!ht]
\centerline{\includegraphics[height=4.8cm]{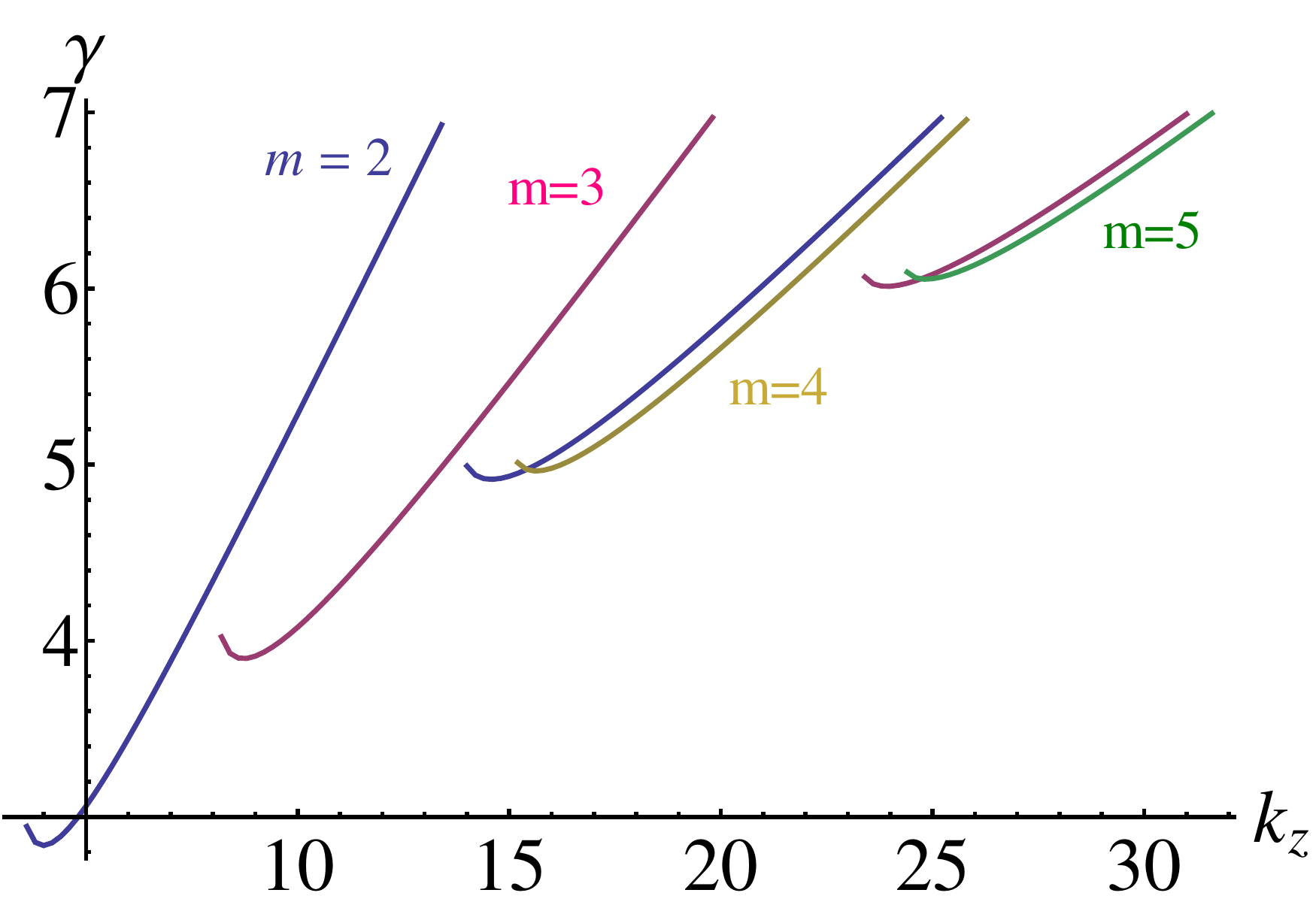}\quad
\includegraphics[height=4.8cm]{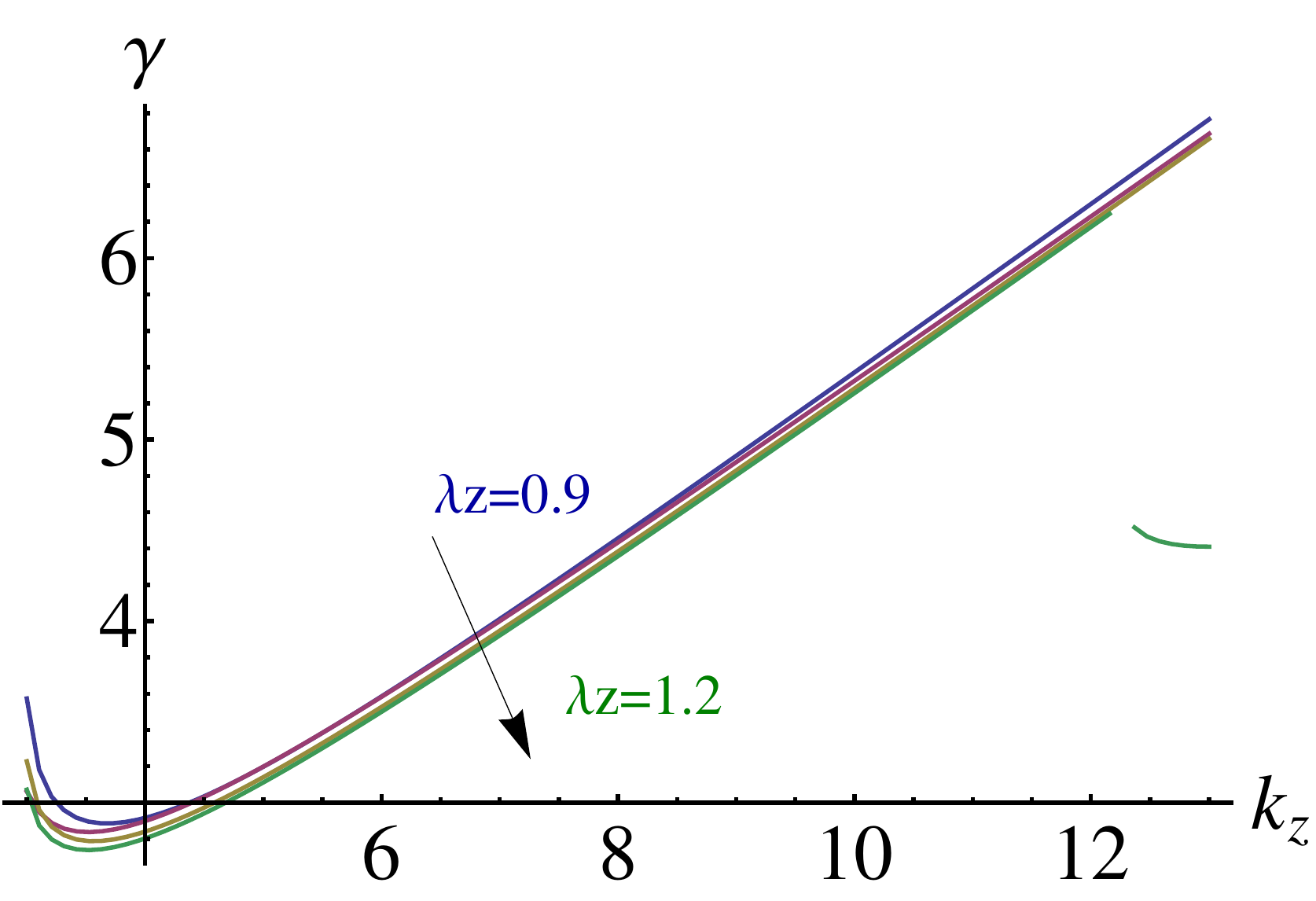}}
\caption{Instability curves for neo-Hookean materials, showing the
critical torsion rate $\gamma$ versus the longitudinal wavenumber
$k_z$, setting $r_o$ as unit length. The curves are depicted at
varying circumferential mode numbers $m$ (left, with $\lambda_z=1$)
and varying axial stretch $\lambda_z$ (right, with $m=2$).
\textbf{The axes origin is set at $\gamma=3$ and $k_z=5$ for matters
of graphic convenience.} } \label{fig1-nh}
\end{figure}

Figure \ref{fig1-nh} depicts our numerical results for the torsion instability of a neo-Hookean
material with $m\geq2$, for
different circumferential modes numbers $m$ and axial pre-stretches $\lambda_z$.

We plot the bifurcation curves with $\gamma r_o$ on the vertical axis and $k_z r_o$ on the horizontal axis.
Hence the first quantity is a measure of the angle of torsion, and the second a measure of the cylinder stubbiness $R_o/L$, according to \eqref{axial-number}.
We find that torsion instability occurs with increasing modes numbers $m$ as the stubbiness increases, and that it
is slightly promoted by an axial extension of the cylinder (and slightly retarded by axial compression).
For instance in simple torsion, i.e. $\lambda_z=1$, the earliest critical threshold of torsion instability is found at $m=2$, $\gamma r_o\simeq2.83743$, for
$k_z r_o\simeq3.9$.


\subsection{Torsion instabilities for a Mooney-Rivlin material}

\begin{figure}[!ht]
\centerline{\includegraphics[height=4.8cm]{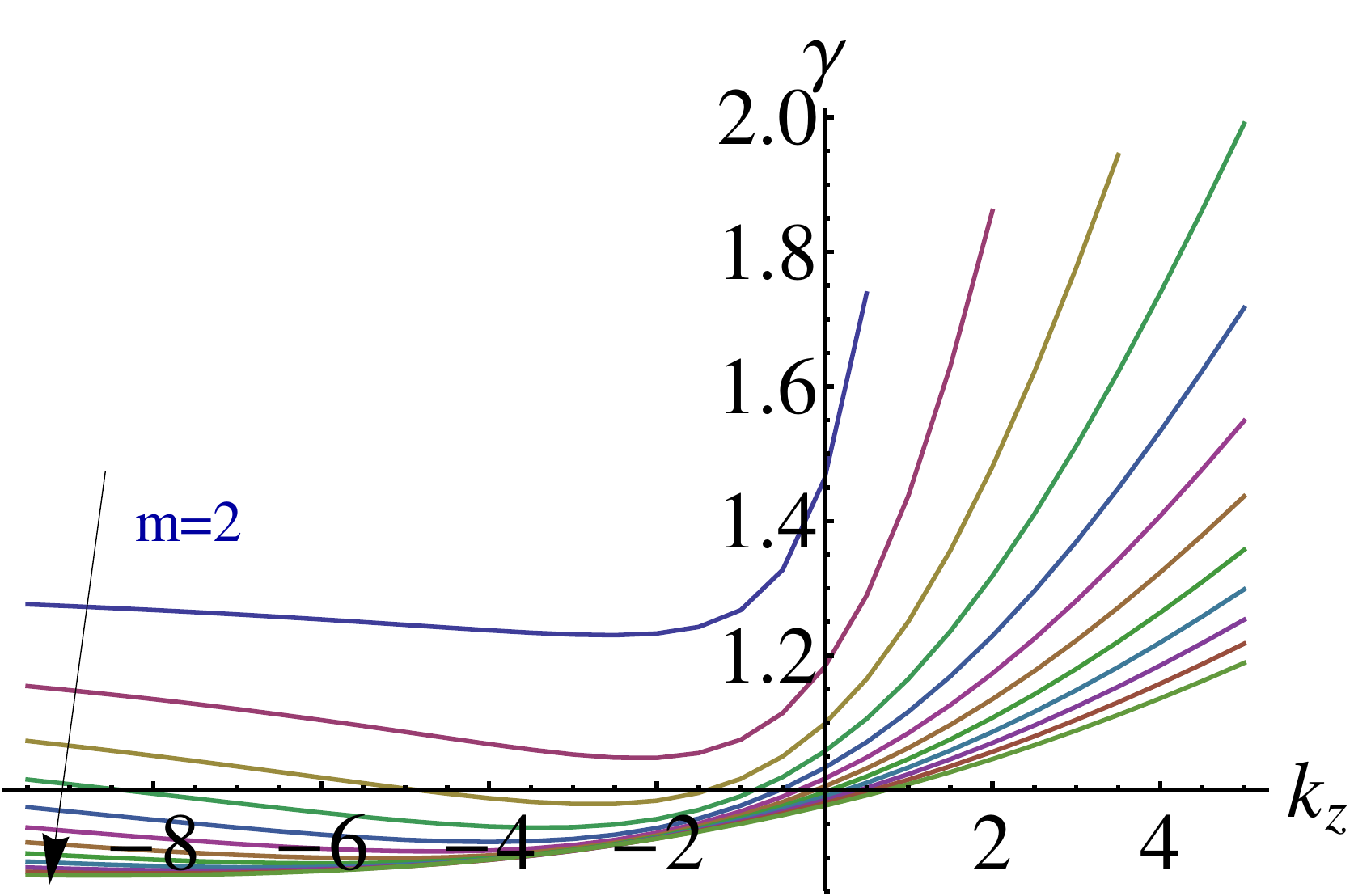}\quad
\includegraphics[height=4.8cm]{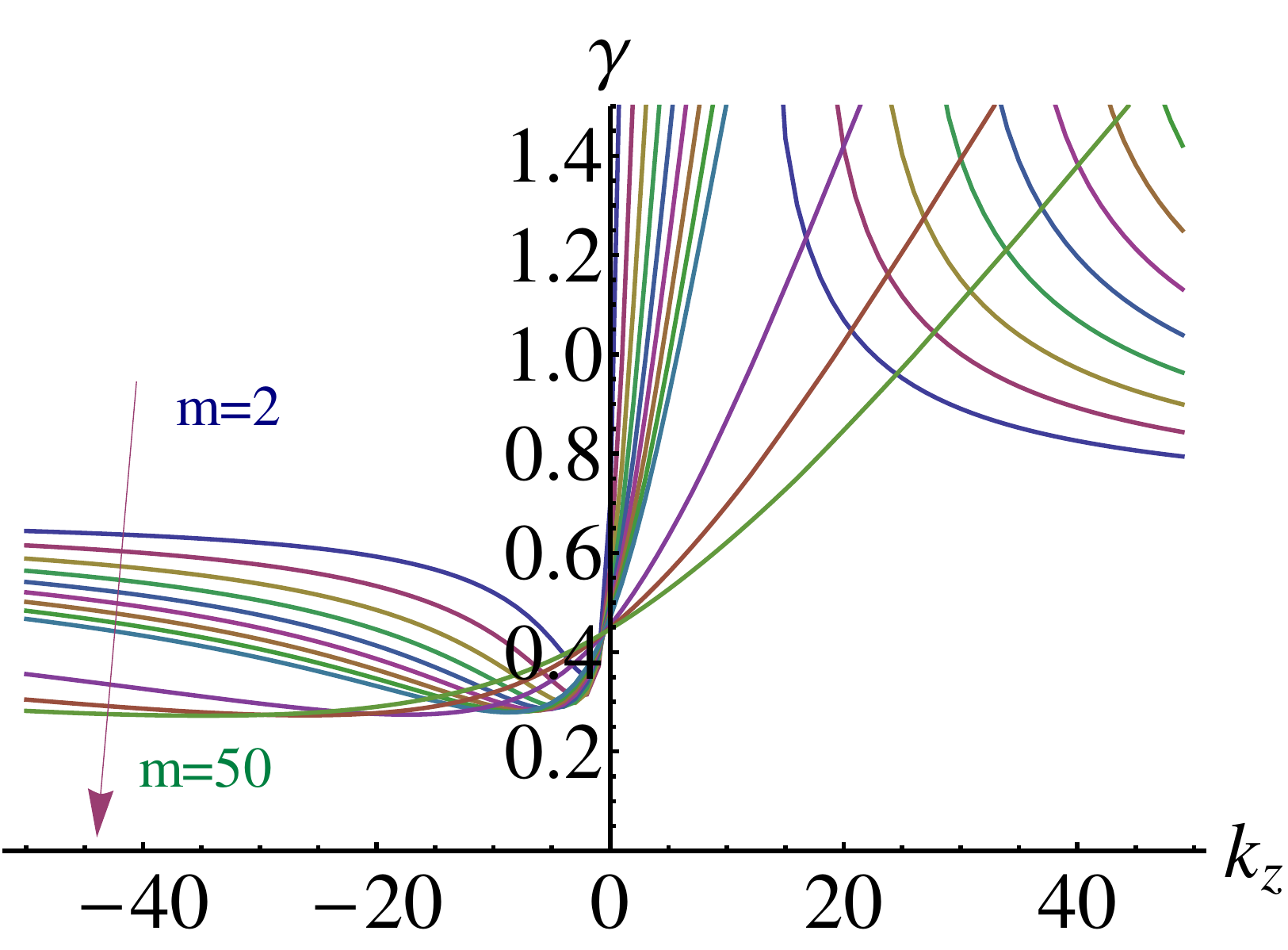}}
\caption{Instability curves for Mooney-Rivlin materials, showing the
critical torsion rate $\gamma$ versus the longitudinal wavenumber
$k_z$, setting $r_o$ as unit length. The curves are depicted at
varying circumferential mode numbers $m$ with $\lambda_z=1$ for $c_2/c_1=1$
(left) and $c_2/c_1=1.5$ (right).} \label{fig2-mr}
\end{figure}
\begin{figure}[!ht]
\centerline{\includegraphics[height=5.2cm]{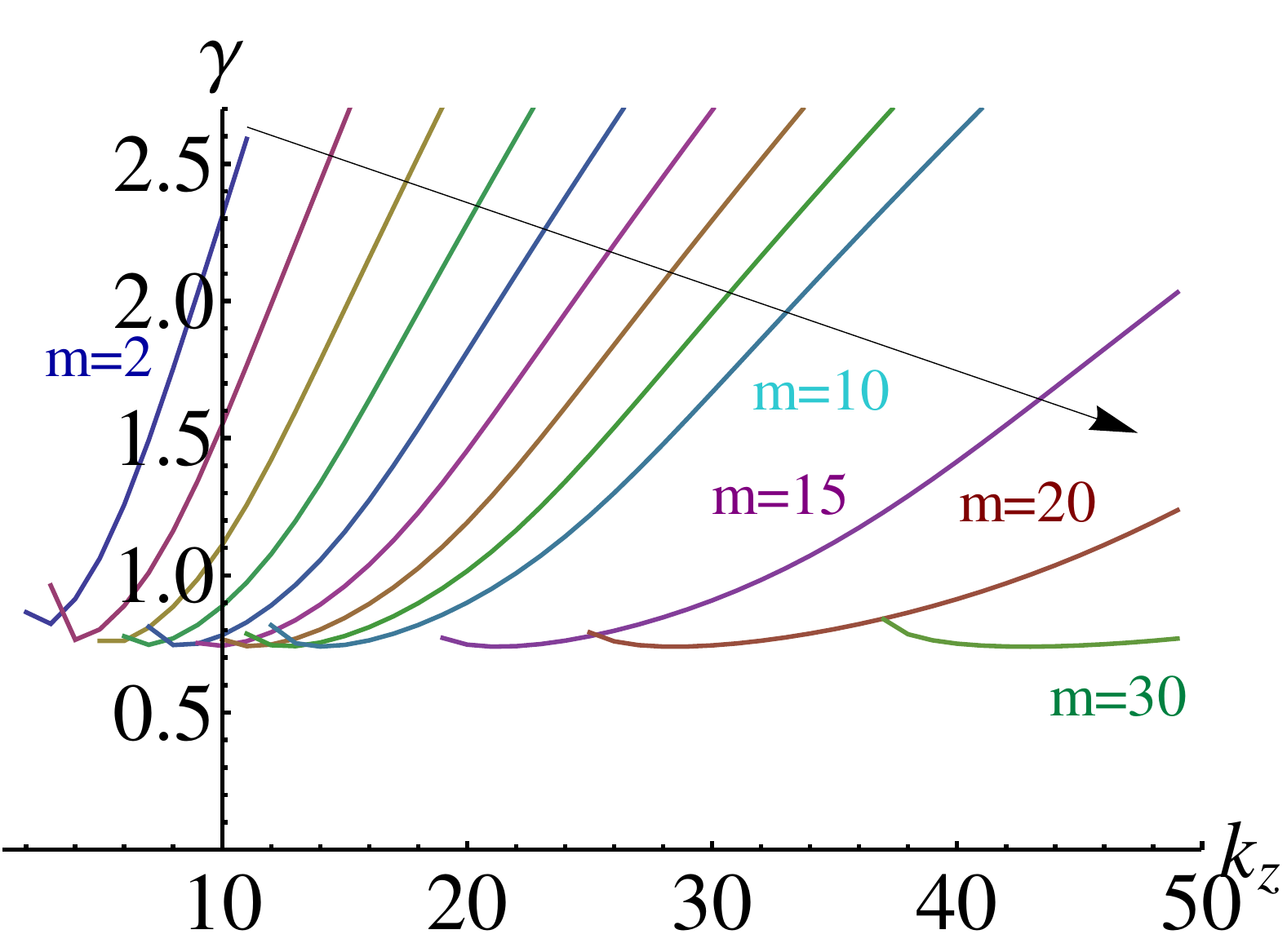}\quad
\includegraphics[height=5.2cm]{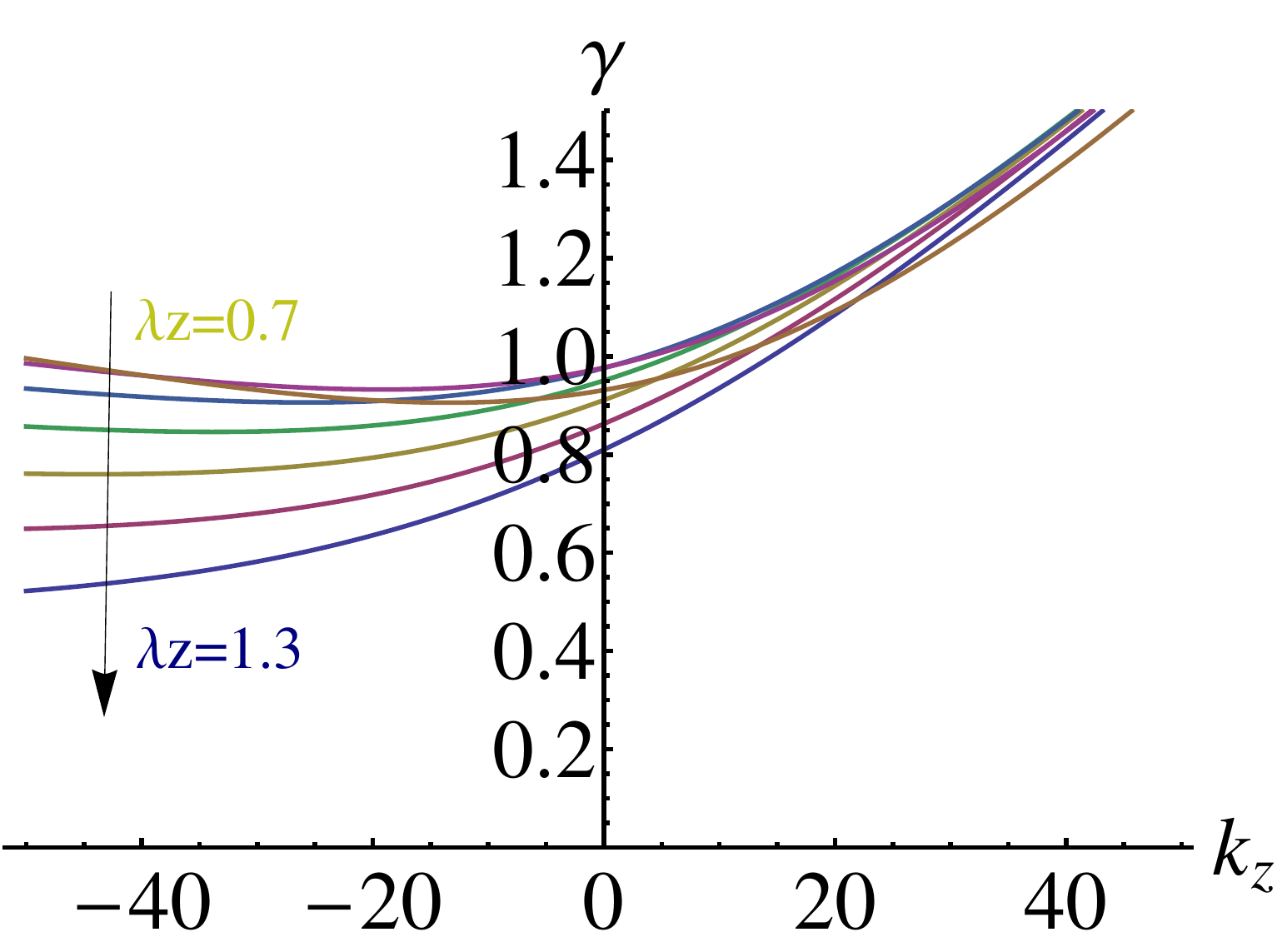}}
\caption{Instability curves for Mooney-Rivlin materials, showing the critical torsion rate
$\gamma$ versus the longitudinal wavenumber $k_z$, setting $R_o$ as
unit length. The curves are depicted at varying circumferential mode numbers
$m$ (left,  where $c_1=0$ and $\lambda_z=1$) and axial stretch
$\lambda_z$ (right, where $c_2/c_1=1$ and $m=50$).}
\label{fig3-mr}
\end{figure}

In the case of a Mooney-Rivlin material ($c_1 \neq 0$,
$c_2\neq0$), we find that the onset of the instability is
strongly dependent on the value of the ratio of the constitutive parameters $c_2/c_1$.
The results of the incremental boundary value problem are collected
in Figures (\ref{fig2-mr}-\ref{fig3-mr}).


In particular, we find that the threshold of the torsion rate
necessary for the onset of the instability is always lower than for
the one found for a neo-Hookean material.
Also, both the
circumferential and the longitudinal modes of the instability
strongly depend on the material constants (e.g. we find $\gamma
r_o\simeq0.2716$, $m=50$ and $k_z r_o\simeq-43.45 $ for
$c_2/c_1=1.5$ in Figure \ref{fig2-mr}).
Although a general trend cannot be clearly identified, it seems that a surface instability mechanism at high mode number $m$ is the dominant scenario for Mooney-Rivlin materials.


\section{Discussion and Conclusions}


We investigated the occurrence of elastic instabilities of a soft incompressible cylinder subjected to a combination of finite axial stretching and finite torsion.
We assumed a Mooney-Rivlin constitutive relation in Eq.\eqref{const} in order to account for weakly nonlinear effects up
to third order in the strain.
The basic axial-symmetric deformation is given by Eq.(\ref{kin}), and the elastic solution for the normal force and
the torque at the top surfaces is given by Eqs. (\ref{N}, \ref{M}).
Using the theory of incremental elastic deformations superimposed on
finite strains, we derived the Stroh formulation of the incremental boundary elastic problem in Eq.\eqref{stroh}.
Introducing the surface impedance matrix ${\bf Z}$ in
Eq.\eqref{impZ}, we rewrote the differential system as a
differential Riccati equation in Eq.\eqref{RiccZ}. In this
theoretical framework, the central-impedance matrix ${\bf Z}_0$
plays a fundamental role for determining the incremental solution.
In order to define a robust numerical procedure, we identified such a matrix as the stable solution of the algebraic Riccati
equation in Eq.(\ref{Riccdis}), hence avoiding singularities at $r=0$.
Finally we computed the numerical solutions by performing iterations on the order parameters driving the elastic bifurcation, whilst the onset of the instability is given by
the target condition of Eq. \eqref{bcZ}.

Leaving aside the oft-studied case of twisting instability, see Figure 1, we focused instead on the occurrence of torsion instability, i.e. the formation of surface wrinkles occurring when the axis displacements are
prevented by geometrical constraints, see Figure 2.
In the case of a neo-Hookean material, the marginal stability curves are depicted in
Figure \ref{fig1-nh}. The critical threshold for the torsion rate in
the case of simple torsion is calculated for $\gamma
r_o\simeq2.83743$, within the experimental range found by
\citet{mora11}.
The marginal stability curves for a Mooney-Rivlin
material are depicted in Figures (\ref{fig2-mr}-\ref{fig3-mr}) for
different ratios $c_2/c_1$ of elastic coefficients  and axial
pre-stretch $\lambda_z$. In particular, we find that a surface
instability mechanism arises, i.e. formation of wrinkles with short
circumferential and axial wavelengths, with lower values of the
critical threshold for the
torsion rate.

In conclusion, we demonstrated that a subclass of torsional
instabilities occur when the axis displacement of the cylinder is
prevented.
The validity of our new theoretical predictions could be easily checked experimentally by applying finite torsion and axial
stretch  with a rheometer to a soft, incompressible cylinder having a  small axial
length/external diameter ratio.


\appendix
\renewcommand{\thesection}{A}
\section*{Appendix A. Stroh matrix for a neo-Hookean material}


Let us consider a neo-Hookean material, setting $c_1=\mu$ and
$c_2=0$ in Eq.\eqref{const}. In this case, from Eqs.(\ref{L}) the
instantaneous elastic moduli reduce to
\begin{equation}
L_{jilk}= \mu b_{jl} \delta_{ik} \label{Lnh}
\end{equation}
Substituting Eqs.(\ref{b},\ref{Lnh}) into Eq.\eqref{stroh}, the
blocks of the Stroh matrix take the following forms:
\begin{equation}
{\bf G}_1= \left[\begin{array}{ccc} \ii & \ii  m &- \ii  k_z r\\[4pt]
-\frac{1}{2} \ii  m \left[2+\gamma^2 \lambda_z^2 \left(R_o^2 - r^2 \lambda_z\right)\right] & -\frac{1}{2} \ii  \left[2+\gamma^2 \lambda_z^2 \left(R_o^2-r^2 \lambda_z\right)\right] & 0 \\[4pt]
\frac{1}{2} \ii  {k_z} r \left[2+\gamma^2 \lambda_z^2 \left(R_o^2 - r^2 \lambda_z\right)\right]&0&0
\end{array}\right],
\label{G1nh}
\end{equation}
and
\begin{equation}
 {\bf G}_2= \left[\begin{array}{ccc}0&0&0\\
0& -\lambda_z/ \mu &0\\ 0&0&  -\lambda_z/\mu \end{array}\right], \qquad
{\bf G}_3= \left[\begin{array}{ccc}
-\dfrac{\alpha_1 \mu }{4 \lambda_z} & -\dfrac{\alpha_{12} \mu }{4 \lambda_z} &
-\dfrac{\alpha_{13} \mu}{2 \lambda_z} \\[10pt]
-\dfrac{\alpha_{12} \mu }{4 \lambda_z} & -\dfrac{\alpha_2 \mu }{4 \lambda_z} & -\dfrac{\alpha_{23} \mu}{\lambda_z} \\[10pt]
-\dfrac{\alpha_{13} \mu }{2 \lambda_z}& -\dfrac{\alpha_{23} \mu}{\lambda_z}& \dfrac{\alpha_3 \mu}{\lambda_z}
\end{array}\right].
\label{G3nh}
\end{equation}
with:
\begin{equation}
\left\{ \begin{array}{lll}& \alpha_1 =\left(-16+8 \gamma k_z m r^2
\lambda_z^3+\gamma^4 m^2 \lambda_z^4 \left(R_o^2-r^2
\lambda_z\right)^2-4 \gamma^2 \lambda_z^2 \left(R_o^2-m^2 R_o^2+2
m^2 r^2 \lambda_z\right)\right.\\ & \qquad \left.+k_z^2 r^2
\left(4+\lambda_z^2 \left(-4 \lambda_z+\gamma^2 \left(R_o^2-r^2
\lambda_z\right) \left(4+\gamma^2 \lambda_z^2 \left(R_o^2-r^2
\lambda_z\right)\right)\right)\right)\right);\\
& \alpha_2=\left(8 \gamma {k_z} m r^2 \lambda_z^3-4 m^2
\left(4+\gamma^2R_o^2 \lambda_z^2\right)+\lambda_z^2 \left(-4 k_z^2
r^2 \lambda_z+4 \gamma^2 \left(R_o^2-2 r^2 \lambda_z\right)+\gamma^4
\lambda_z^2
\left(R_o^2-r^2 \lambda_z\right)^2\right)\right);\\
& \alpha_3=\left(m^2+\gamma r^2 \left(-2 k_z m+\gamma m^2-\gamma
k_z^2 r^2\right) \lambda_z^3+k_z^2 r^2 \left(3+\gamma^2 R_o^2
\lambda_z^2+\lambda_z^3\right)\right);\\
& \alpha_{12}=\left(8 \gamma k_z r^2 \lambda_z^3+m \left(-16-8
\gamma^2 r^2 \lambda_z^3+\gamma^4 \lambda_z^4 \left(R_o^2-r^2
\lambda_z\right)^2\right)\right);\\
& \alpha_{13}=k_z r \left(4+\gamma^2 \lambda_z^2 \left(R_o^2-r^2
\lambda_z\right)\right)\\& \alpha_{23}=k_z m r \left(3+\gamma^2
\lambda_z^2 \left(R_o^2-r^2 \lambda_z\right)\right).
\end{array} \right.
\end{equation}


\section*{Acknowledgements}


Partial funding by the European Community grant ERG-256605 (FP7
program), the INSERM grant OTPJ12U170 (Plan Cancer), and by a ``New
Foundations'' award from the Irish Research Council are gratefully
acknowledged by the first and the second author, respectively. {The
authors also thank Fionnan O'Reilly (Dublin) and Michael Gilchrist
(Dublin) for technical assistance, and an anonymous referee for
constructive comments on an earlier version of the paper.}



\end{document}